\pdfoutput=1

\documentclass[11pt,twoside,a4paper,cmspaper,final,collab]{cms-tdr}
\def\svnVersion{d585f2d-D}\def\svnDate{2019/09/02}\def\cmsCernNoTag{CERN-EP-2018-344}\def\cmsCernDate{\today}\def\cmsMessage{"Published in the European Physical Journal C as \href{http://dx.doi.org/10.1140/epjc/s10052-019-7276-4}{\doi{10.1140/epjc/s10052-019-7276-4}.}"}

\begin{document}\cmsNoteHeader{SMP-17-009}

\hyphenation{had-ron-i-za-tion}
\hyphenation{cal-or-i-me-ter}
\hyphenation{de-vices}
\RCS$HeadURL$
\RCS$Id$
\newlength\cmsFigWidth
\ifthenelse{\boolean{cms@external}}{\setlength\cmsFigWidth{0.85\columnwidth}}{\setlength\cmsFigWidth{0.4\textwidth}}
\ifthenelse{\boolean{cms@external}}{\providecommand{\cmsLeft}{top\xspace}}{\providecommand{\cmsLeft}{left\xspace}}
\ifthenelse{\boolean{cms@external}}{\providecommand{\cmsRight}{bottom\xspace}}{\providecommand{\cmsRight}{right\xspace}}

\newlength\cmsTabSkip\setlength{\cmsTabSkip}{1ex}
\newenvironment{tolerant}[1]{\par\tolerance=#1\relax}{ \par }

\ifthenelse{\boolean{cms@external}}{\providecommand{\cmsTable}[1]{#1}}{\providecommand{\cmsTable}[1]{\resizebox{\textwidth}{!}{#1}}}

\newcommand{\alpsmz}{\ensuremath{\alpS(m_{\Z})}\xspace}
\newcommand{\chisq}{\ensuremath{\chi^2}\xspace}
\newcommand{\chisqndof}{\ensuremath{\chi^2/n_\mathrm{dof}}\xspace}
\newcommand{\NLOJETPP} {{\textsc{NLOJet++}}\xspace}
\newcommand{\fastNLO} {{\textsc{fastNLO}}\xspace}
\newcommand{\PYTHIAE} {{\PYTHIA8}\xspace}
\newcommand{\POWHEGTWOJET} {{\textsc{ph-2j}}\xspace}
\newcommand{\POWHEGTHREEJET} {{\textsc{ph-3j}}\xspace}
\newcommand{\RooUnfold} {{\textsc{RooUnfold}}\xspace}
\newcommand{\TUnfold} {{\textsc{TUnfold}}\xspace}
\newcommand{\ptmax}{\ensuremath{\pt^{\text{max}}}\xspace}
\newcommand{\mur}{\ensuremath{\mu_r}\xspace}
\newcommand{\muf}{\ensuremath{\mu_f}\xspace}
\newcommand{\rbtrr}{\rule[-0.8ex]{0ex}{3.2ex}}
\newcommand{\dphiOneTwo}{\ensuremath{\Delta\phi_{12}}\xspace}

\renewcommand{\floatpagefraction}{.9}%	

\cmsNoteHeader{SMP-17-009}
\title{Azimuthal separation in nearly back-to-back jet topologies in inclusive 2- and 3-jet events in $\Pp\Pp$ collisions at $\sqrt{s}=13\TeV$}
\titlerunning{Azimuthal separation in nearly back-to-back jet topologies  in inclusive 2- and 3-jet events at $\sqrt{s}=13\TeV$}

\date{\today}

\abstract{
          A measurement for inclusive 2- and 3-jet events of the azimuthal correlation between the two jets with the largest
          transverse momenta, \dphiOneTwo , is presented.
          The measurement considers events where the two leading jets are
          nearly collinear (``back-to-back") in the transverse plane and is performed for several ranges
          of the leading jet transverse momentum.
	  Proton-proton collision data
	  collected with the CMS experiment at a center-of-mass energy
	  of $13\TeV$ and corresponding to an integrated luminosity of $35.9\fbinv$ are used.
	  Predictions based on calculations using matrix elements at leading-order and next-to-leading-order accuracy in
	  perturbative quantum chromodynamics supplemented with leading-log parton showers and hadronization are generally in agreement with
	  the measurements. Discrepancies between the measurement and theoretical predictions are as large as 15\%, mainly in the
	  region $177^\circ < \dphiOneTwo < 180^\circ$. The 2- and 3-jet measurements are not simultaneously described by any of models.
}

\hypersetup{
pdfauthor={CMS Collaboration},
pdftitle={Azimuthal angular correlations in back-to-back dijets events},
pdfsubject={CMS},
pdfkeywords={CMS, physics, QCD, jets, azimuthal angle, MC generators}
}

\maketitle

\section{Introduction}
\label{sec:intro}

Collimated streams of particles (jets) can be produced in highly energetic parton-parton interactions in proton-proton (s ) collisions, and their properties
are described by the theory of strong interactions, quantum chromodynamics (QCD). In  the lowest order perturbative QCD (pQCD), two jets with high transverse momenta \pt\ are produced ``back-to-back" in the transverse plane. Higher order corrections lead to deviations from this configuration.
Experimentally, this can be investigated by the measurement of
the azimuthal separation, $\dphiOneTwo=\abs{\phi_\text{jet1}-\phi_\text{jet2}}$, between the two leading \pt jets in the transverse plane.
Within the framework of pQCD, a final state with three or more partons is required for significant deviations from $\dphiOneTwo=180^\circ$.
However, when deviations of \dphiOneTwo from $180^\circ$ are small,
a pQCD calculation at a fixed order in the strong coupling \alpS\
becomes unstable and a resummation of soft parton emissions to all orders in \alpS\ has to be performed. This resummation is approximated through the use of parton showers in Monte Carlo (MC) event generators.

Azimuthal correlations in inclusive 2-jet events have been measured previously by the D0 Collaboration
in $\Pp\PAp$ collisions at a center-of-mass energy of $\sqrt{s}=1.96\TeV$ \cite{bib:D0_1,bib:D0_2}, in $\Pp\Pp$ collisions
by the ATLAS Collaboration at $\sqrt{s}=7\TeV$  \cite{bib:ATLAS},
and by the CMS Collaboration at $\sqrt{s}=7$, $8$, and $13\TeV$  \cite{bib:CMS,bib:CMS_2,bib:CMS_2_1}, but none of the measurements considered in detail the region close to the back-to-back configuration.
A detailed study of azimuthal correlations close to the back-to-back configuration allows a more precise test of different resummation strategies, and it is a first step towards an improved understanding of the effects of soft initial and final state gluons \cite{Catani:2017tuc,Catani:2014qha}. The leading- and next-to-leading-logarithm contributions to the dijet azimuthal angular correlation have been investigated in \cite{Banfi:2008qs,Sun:2014gfa,Sun:2015doa}. The effects of applying a transverse momentum dependent parton showering to the dijet azimuthal angular correlation were studied in \cite{Hautmann:2008vd}.

In this article measurements are reported of the normalized inclusive 2-jet distribution as a function of the azimuthal   separation $\dphiOneTwo$ between the two leading \pt jets (jets~1 and~2),
 \begin{equation}
  \frac{1}{\sigma_{\ptmax}}
  \frac{\rd\sigma}{\rd\dphiOneTwo},
  \label{eq:2J}
\end{equation}
in several intervals of the leading jet \pt (\ptmax ) within the rapidity range $\abs{y}<2.5$.
The total dijet cross section $\sigma_{\ptmax}$ is measured within each range of \ptmax integrated over the full range in \dphiOneTwo.
The binning of the measurement presented here is much finer than that of Ref.~\cite{bib:CMS_2_1}.
We consider $\dphiOneTwo$ in the range $170^{\circ} < \dphiOneTwo \leq 180^{\circ}$.

The inclusive 3-jet distributions, differential in \dphiOneTwo and \ptmax, with the \pt of third highest \pt jet typically being 1-2 orders of magnitude smaller than \ptmax, are also suitable to test resummation effects arising from the presence of multiple scales in the interaction.
Measurements of the inclusive 3-jet distribution normalized to $\sigma_{\ptmax}$ are also presented, for several ranges of \ptmax, and within $\abs{y}<2.5$.

The measurements are performed using data collected from \Pp\Pp\ collisions at $\sqrt{s}=13\TeV$ during 2016 with the CMS experiment at the CERN LHC, corresponding to an integrated luminosity of 35.9\fbinv.

\section{The CMS Detector}
\label{sec:detector}

The central feature of the CMS detector is a superconducting solenoid, 13\unit{m} in
length and 6\unit{m} in inner diameter, providing an axial magnetic field of 3.8\unit{T}.
Within the solenoid volume are a silicon pixel and strip tracker,
a lead tungstate crystal electromagnetic calorimeter (ECAL)
and a brass and scintillator hadron calorimeter (HCAL), each composed
of a barrel and two endcap sections.
Charged-particle trajectories are measured by the tracker
with full azimuthal coverage within pseudorapidities $\abs{\eta}<2.5$.
The ECAL, which is equipped with a preshower detector in the endcaps,
and the HCAL cover the region $\abs{\eta}<3.0$.
Forward calorimeters extend the pseudorapidity coverage provided
by the barrel and endcap detectors to the region $3.0<\abs{\eta}<5.2$.
Finally, muons are measured up to $\abs{\eta}<2.4$ by gas-ionization
detectors embedded in the steel flux-return yoke outside the solenoid.
A detailed description of the CMS detector together with a definition of the coordinate system
used and the relevant kinematic variables can be found in Ref.~\cite{Chatrchyan:2008aa}.

\section{Theoretical predictions}
\label{sec:theory}

Simulations from leading-order (LO) and next-to-LO (NLO) MC event generators are investigated. Among the LO event generators, both \PYTHIAE \cite{bib:pythia8} (version 8.219) and  \HERWIGpp \cite{bib:herwigpp} (version 2.7.1) are used for predictions
because they feature different  parton showering (PS) algorithms for soft and collinear parton radiation at leading-log accuracy.
In \PYTHIAE the PS emissions cover a region of phase space ordered in $x$ (fraction of the proton momentum carried by the parton) and the \pt of the emitted parton, whereas in \HERWIGpp the parton emissions are ordered in $x$ and the angle of the radiated parton (angular ordering). The Lund string model~\cite{Andersson:1998tv} is used for hadronization in
\PYTHIAE~\cite{bib:pythia8}, whereas  in \HERWIGpp the cluster fragmentation model \cite{Webber:1983if} is applied.
Multiparton interactions (MPI) are simulated in \PYTHIAE\ (tune CUETP8M1~\cite{Khachatryan:2015pea} with the parton distribution function (PDF) set NNPDF2.3LO~\cite{Ball:2013hta,Ball:2011uy}) and  in \HERWIGpp\  (tune CUETHppS1~\cite{Khachatryan:2015pea} with the PDF set CTEQ6L1~\cite{Pumplin:2002vw}) with parameters tuned to measurements in \Pp\Pp\ collisions at the LHC and $\Pp\PAp$ collisions at the Tevatron.

The \MGvATNLO~\cite{bib:madgraph5} version 2.3.3 event generator (labelled as \MADGRAPH in the following) interfaced with
\PYTHIAE with tune CUETP8M1 is also used in the analysis.
Processes with up to 4 final-state partons at LO accuracy are calculated using the NNPDF2.3LO PDF set.
The \kt-MLM matching procedure \cite{Alwall:2007fs} is used with a matching scale of 10\GeV.

Among the NLO event generators, predictions obtained using the \POWHEG\ {\sc box}
library~\cite{bib:Nason:2004rx,bib:Frixione:2007vw,bib:Alioli:2010xd} (version 2) with the PDF set NNPDF3.0NLO~\cite{Ball:2014uwa} are considered. The event generators \PYTHIAE (tune CUETP8M1) and \HERWIGpp (tune CUETHppS1) are used to simulate PS, hadronization, and MPI.
The \POWHEG generator in dijet mode \cite{bib:POWHEG_Dijet}, referred to as \POWHEGTWOJET, provides an NLO dijet calculation, which is accurate to LO for the azimuthal correlation between the leading jets. The \POWHEG generator in three-jet mode~\cite{Kardos:2014dua} (using the MiNLO scheme~\cite{Hamilton:2012rf,Hamilton:2012np}), referred to as \POWHEGTHREEJET, provides an NLO $2\to3$ calculation.
For the \POWHEGTWOJET matrix elements (ME), a minimum \pt of 100 GeV is required on the partons in the Born process, while for the \POWHEGTHREEJET ME the minimum is lowered to 10 GeV to ensure coverage of the full phase space. These thresholds are applied  to optimize the generation of events in the phase space of interest.
The matching between the \POWHEG matrix element calculations and the \PYTHIAE underlying event (UE)~\cite{Khachatryan:2015pea} simulation is performed by using
the shower-veto procedure (UserHook option 2 \cite{bib:pythia8}). The matching between the \POWHEG matrix element calculations and the \HERWIGpp UE~\cite{Khachatryan:2015pea} is performed by using a truncated shower \cite{bib:Nason:2004rx}.

\begin{tolerant}{800}
Events generated by \PYTHIAE (tune CUETP8M1), \HERWIGpp (tune CUETHppS1), and \MADGRAPH interfaced with \PYTHIAE (tune CUETP8M1) are passed through a full detector simulation based on \GEANTfour \cite{bib:geant}. The simulated events events are reconstructed with standard CMS programs.
\end{tolerant}

   Table~\ref{tableMC} summarizes the theoretical predictions used in the present analysis.

\begin{table*}[htbp]
  \centering
    \topcaption{ Monte Carlo event generators, parton densities, and underlying event tunes used for comparison with measurements.}
    \label{tableMC}
\cmsTable{
\begin{tabular}{l l l l}
 {Matrix element generator} &  {Simulated diagrams} &  {PDF set} &  {Tune}\\ \hline
 \\[-1.5ex]
 \PYTHIA{}~8.219~\cite{bib:pythia8} &  2$\to$2 (LO) &  NNPDF2.3LO~\cite{Ball:2013hta,Ball:2011uy} &  CUETP8M1~\cite{Khachatryan:2015pea}
 \\[1.0ex]
\HERWIGpp~2.7.1~\cite{bib:herwigpp} &  2$\to$2 (LO) &  CTEQ6L1~\cite{Pumplin:2002vw} &  CUETHppS1~\cite{Khachatryan:2015pea}
 \\[1.0ex]
\MADGRAPH~\cite{bib:madgraph5,Alwall:2007fs}+  \PYTHIA{}~8.219~\cite{bib:pythia8} &  2$\to$2, 2$\to$3, 2$\to$4 (LO) &  NNPDF2.3LO~\cite{Ball:2013hta,Ball:2011uy} &  CUETP8M1~\cite{Khachatryan:2015pea}
\\[2.5ex]
\POWHEGTWOJET{}~\cite{bib:Frixione:2007vw,bib:Alioli:2010xd,bib:Nason:2004rx}   +  \PYTHIA{}~8.219~\cite{bib:pythia8} &  2$\to$2 (NLO) &  NNPDF3.0NLO~\cite{Ball:2014uwa} &  CUETP8M1~\cite{Khachatryan:2015pea}
\\[2.5ex]
 \POWHEGTWOJET{}~\cite{bib:Frixione:2007vw,bib:Alioli:2010xd,bib:Nason:2004rx} + \HERWIGpp~2.7.1~\cite{bib:herwigpp}   &  2$\to$2 (NLO) &  NNPDF3.0NLO~\cite{Ball:2014uwa} &  CUETHppS1~\cite{Khachatryan:2015pea}
\\[2.5ex]
 \POWHEGTHREEJET{}~\cite{bib:Frixione:2007vw,bib:Alioli:2010xd,bib:Nason:2004rx} +  \PYTHIA{}~8.219~\cite{bib:pythia8}&  2$\to$3 (NLO)&  NNPDF3.0NLO~\cite{Ball:2014uwa} &  CUETP8M1~\cite{Khachatryan:2015pea}
\\[1.0ex]
\hline
\end{tabular}
}
\end{table*}

\section{Jet reconstruction and event selection}
\label{sec:selectionReconstruction}

The measurements are based on data samples collected with single-jet high-level triggers~\cite{CMS_HLT,CMS_trigger}.
The five single-jet triggers require at least one jet in the event with $\pt > 140$, 200, 320, 400, or $450 \GeV$ within the full rapidity coverage of the CMS calorimetry.
Table~\ref{table:turn_ons} shows  the
various \ptmax regions accessed by the various triggers and the integrated luminosity for each trigger
     in the analysis. Each trigger is fully efficient for jets in the corresponding \pt range in Table~\ref{table:turn_ons}.

\begin{table*}[htb!]
  \centering
    \topcaption{The integrated luminosity for each trigger
      sample in the analysis, and trigger used for each \ptmax range.  \label{table:turn_ons}
}
    \begin{tabular}{lccccc}
      HLT \pt threshold (\GeVns{})   &  140    &  200    &  320  & 400  &  450 \\
      \hline
     \lumi ($\text{fb}^{-1}$)\rbtrr &  0.024  &  0.11  &  1.77 &  5.2 & 36 \\
     \ptmax region (\GeVns{})   &  200--300   & 300--400    &  400--500  & 500--600  &  $>$600 \\
      \hline
    \end{tabular}
\end{table*}

\begin{tolerant}{1200}
Particles are reconstructed and identified using a particle-flow (PF) algorithm~\cite{bib:parflow},
which utilizes an optimized combination of information from the various elements of the CMS detector.
Jets are reconstructed by clustering the four-vectors of the PF candidates with the infrared-
and collinear-safe anti-\kt clustering algorithm~\cite{bib:antikt} with a distance parameter $R=0.4$.
The clustering is performed with the \FASTJET package~\cite{bib:fastjet}.
To reduce the contribution to the reconstructed jets from additional \Pp\Pp\  interactions within the same bunch crossing (pileup), the charged-hadron subtraction technique~\cite{bib:jes8} is used  to remove tracks identified as originating from pileup vertices.
The average number of pileup interactions per single bunch crossing observed in the data is about 27.
 The pileup contribution from neutral hadrons is corrected using   a jet-area-based correction technique~\cite{Cacciari:2007fd}.
\end{tolerant}

For this analysis, jets with
rapidity $\abs{y}< 5.0$ are reconstructed. For both the inclusive 2- and 3-jet samples, the events are selected by requiring the two highest \pt\ jets to have  $\abs{y}< 2.5$ and $\pt > 100\GeV$. For the inclusive 3-jet events a third jet with  $\pt >30\GeV$ and $\abs{y}< 2.5$ is required. Contributions from pileup are negligible because the pileup removal algorithm has an efficiency of $\sim$99\% for jets with $30 <\pt < 50\GeV$ and $\abs{y}< 2.5$~\cite {CMS-PAS-JME-13-005}.

\section{Measurements of the normalized inclusive 2- and 3-jet distributions}
\label{sec:measurement}

The normalized inclusive 2- and 3-jet distributions as a function of \dphiOneTwo are corrected for detector resolution.
We achieve this by unfolding the observables to the level of stable final-state particles. In this way, a direct comparison of these measurements to results from other experiments
and to QCD predictions is possible.
Particles are considered stable if their mean decay length is larger than 1\unit{cm}.

The unfolding procedure is based on the D'Agostini algorithm~\cite{DAgostini:1994zf}, which is  implemented in the
\RooUnfold\ package~\cite{bib:RooUnfold}, by using a response matrix that maps the generated jets onto the jets reconstructed by the CMS detector. The regularization (number of iterations) of the unfolding procedure is chosen by comparing the difference in \chisq between data and MC at detector level to that between data and MC at particle level. The consistency of the unfolding procedure is checked against the alternative \TUnfold\ package~\cite{Schmitt:2012kp,Schmitt:2016orm}, which uses a least square minimization with Tikhonov regularization. Both methods provide equivalent results.
The unfolding is performed in \dphiOneTwo.
The response matrices are obtained using simulated events from the \PYTHIAE event generator with the tune CUETP8M1.
The difference between the unfolded distributions and the distributions at detector level range from $\sim$1\% for the low \ptmax regions up to $\sim$5\% for the high \ptmax regions.

The sources of systematic uncertainties arise primarily from the jet energy scale calibration (JES), the  jet energy resolution (JER), the \dphiOneTwo resolution, and the model dependence of the unfolding matrix. The effect of migrations between \ptmax regions is very small because of the normalization of the cross sections in each \ptmax range and therefore is neglected.

The \dphiOneTwo\ resolution is ${\sim}0.5^{\circ}$, as obtained from fully simulated event samples from  \PYTHIAE and \MADGRAPH.
A bin size of $1^{\circ}$ is  a compromise between the ability to study the back-to-back region and the impact of the unfolding correction of $\sim$2\%. In Ref.~\cite{bib:CMS_2_1} the study is focused on a different \dphiOneTwo\ region, and a coarser bin size is chosen to account for the smaller size of the data sample.

Alternative response matrices are obtained by using the \dphiOneTwo\ resolution determined from fully simulated events. This resolution is varied by $\pm$10\%, an amount that is motivated by the observed difference between data and simulation. The resulting uncertainty is estimated to be below 1\%.

An additional systematic uncertainty is caused by the dependence of the response matrix
on the choice of the MC generator. Alternative response matrices are built using the \HERWIGpp
and \MADGRAPH + \PYTHIAE event generators.
Because this analysis uses a finer binning compared with that of Ref.~\cite{bib:CMS_2_1}, the sensitivity to the uncertainty in the unfolding is increased. The observed effect from bin migration is less than 2\%.

The JER and shifts in the JES can cause events to migrate between the \ptmax regions. The JES uncertainties on the energy measurement are estimated to be 1--2\%
\cite{bib:jes8}.
The resulting JES uncertainties in the normalized inclusive 2-jet distributions due to bin migrations are less than 2\%,
whereas for the normalized inclusive 3-jet distributions they are less than 3\%.
The effect of the JER uncertainties \cite{bib:jes8} is estimated by varying the JER parameters by one standard deviation up and down and comparing the results before and after the changes.
The JER-induced uncertainties are less than 0.2\% for the inclusive 2-jet \dphiOneTwo\ measurement and below 0.4\% for the normalized inclusive 3-jet measurement.

\section{Comparison to theoretical predictions}
\label{sec:compMC}
In this section the measurements are compared with different theoretical predictions introduced in Section~\ref{sec:theory}. In all figures displaying ratios, the solid band indicates the total experimental uncertainty and the error bars represent the statistical uncertainties from the simulation.
In the figures displaying the normalized distributions, the error bars on the data represent the total experimental uncertainty and the error bars on the predictions represent the statistical uncertainty of
the simulation. The uncertainties are often so small that the bars are not visible.

\begin{figure*}[hbt]
  \centering
  \includegraphics[width=0.4\textwidth]{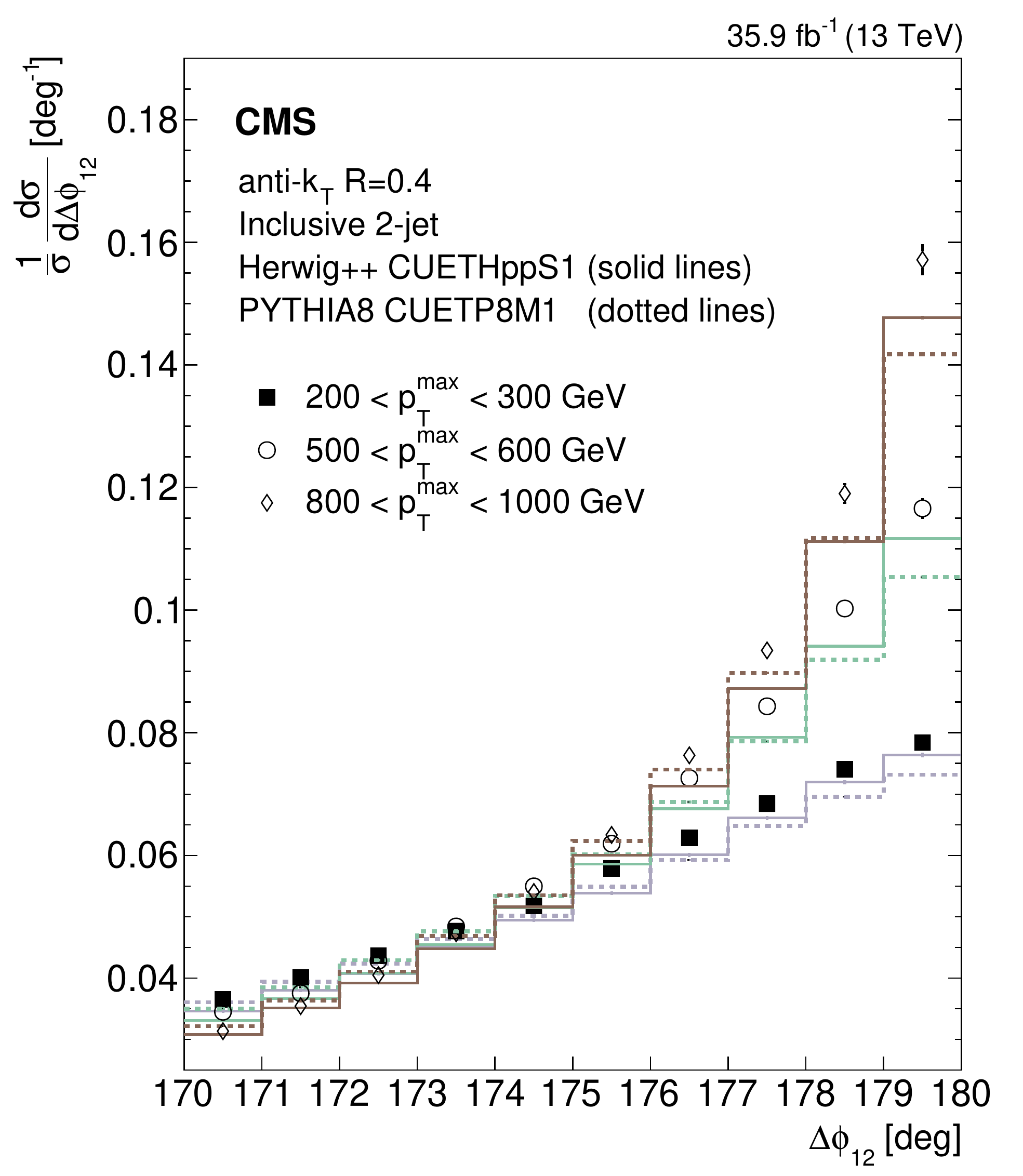}
  \includegraphics[width=0.4\textwidth]{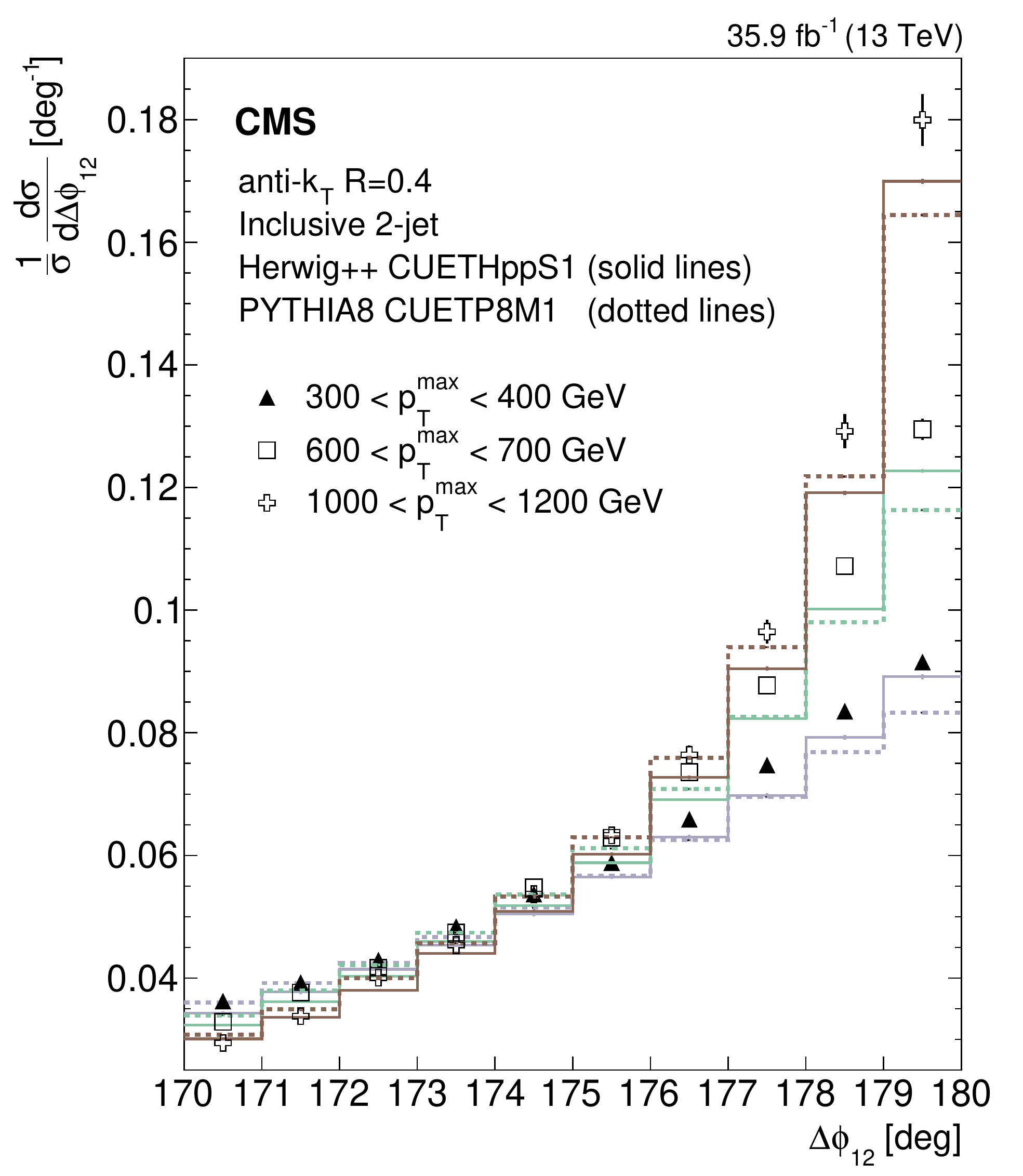}
  \caption{Normalized inclusive 2-jet distributions as a function of the azimuthal separation of the two leading jets \dphiOneTwo
    for different \ptmax regions.
The data are represented by the markers and the theory by histograms. Overlaid with the data are predictions from the \HERWIGpp event generator (solid lines) and \PYTHIAE (dotted lines). The total experimental uncertainty is depicted as error bars on the measurement.}
  \label{fig:2J_particle_xsection_MC_data}
\end{figure*}

The unfolded normalized inclusive 2-jet distribution as a function of  \dphiOneTwo is shown in
Fig.~\ref{fig:2J_particle_xsection_MC_data}, and compared with the predictions from \HERWIGpp (solid lines ) and \PYTHIAE (dotted lines) for different \ptmax regions.
The distributions are strongly peaked at $180^{\circ}$ and become steeper with increasing \ptmax.
\begin{figure*}[hbt]
  \centering
  \includegraphics[width=0.8\textwidth]{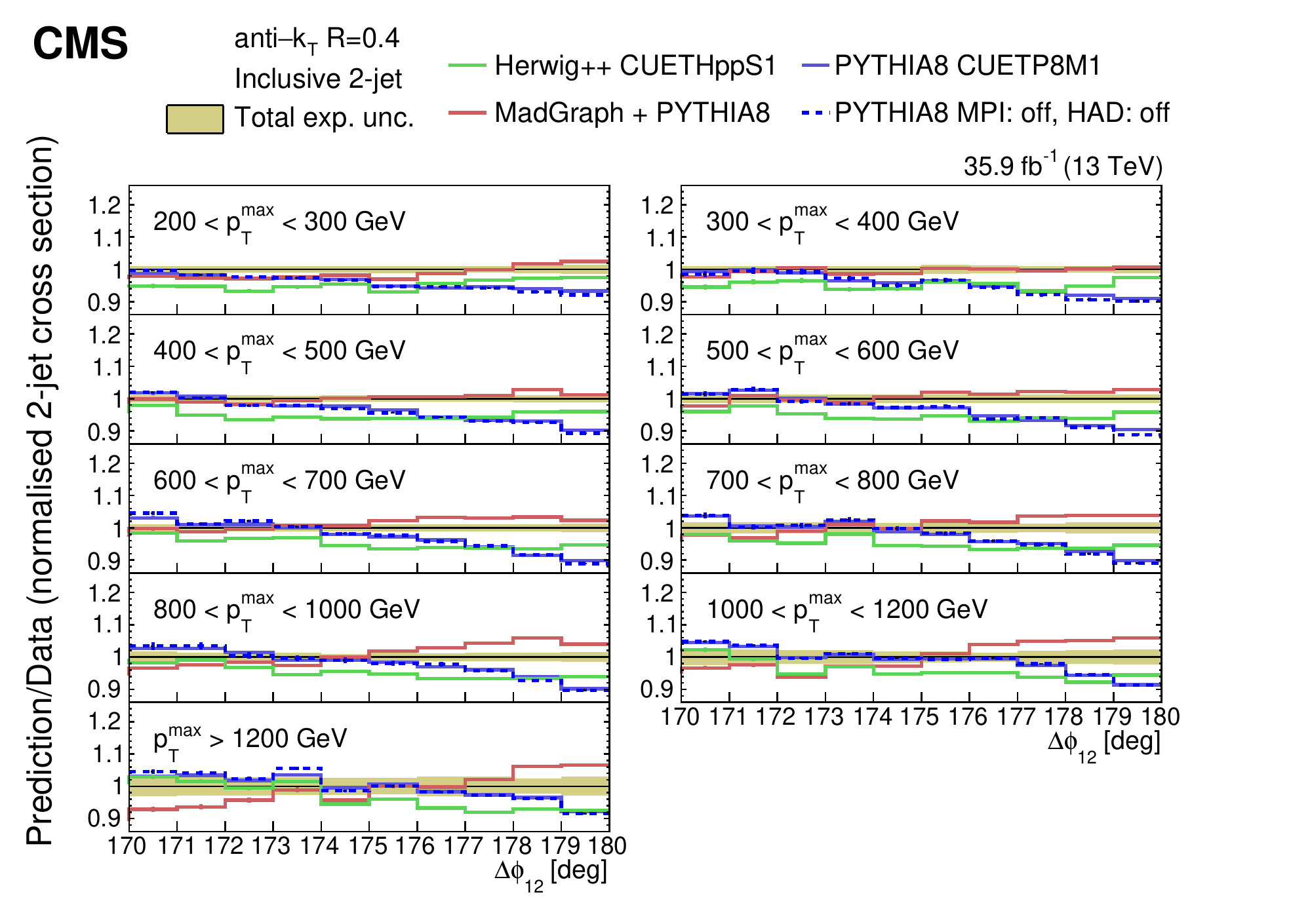}
  \caption{Ratios of the normalized
    inclusive 2-jet distributions for the \PYTHIAE, \HERWIGpp, and \MADGRAPH + \PYTHIAE predictions to data  as a function of the azimuthal   separation of the two leading jets \dphiOneTwo, for all the \ptmax regions.
    The solid band indicates the total experimental uncertainty and
    the error bars on the MC points represent the statistical
    uncertainty of the simulated data.}
  \label{fig:2J_LO_data}
\end{figure*}
The ratio  of the \PYTHIAE, \HERWIGpp, and \MADGRAPH + \PYTHIAE event generator predictions to data are depicted in Fig.~\ref{fig:2J_LO_data} for
the inclusive 2-jet distributions in the nine \ptmax ranges.
Among the event generators, \PYTHIAE and \HERWIGpp show the largest deviations from the measurements for the $\ptmax < 800$ GeV regions in the inclusive 2-jet case, and the \MADGRAPH + \PYTHIAE event generator gives the best description in the same regions. The three generators show large deviations from the measurements in the  $\ptmax > 800$ GeV regions.
The nonperturbative corrections are estimated to be small (below 1.5\%) by comparing the predictions from \PYTHIAE without the simulation of multi-parton interactions and hadronization (dashed blue curve) to the predictions from \PYTHIAE when these effects are included (solid blue curve). The nonperturbative correction factors are available in HepData.

\begin{figure*}[htb]
  \centering
  \includegraphics[width=0.8\textwidth]{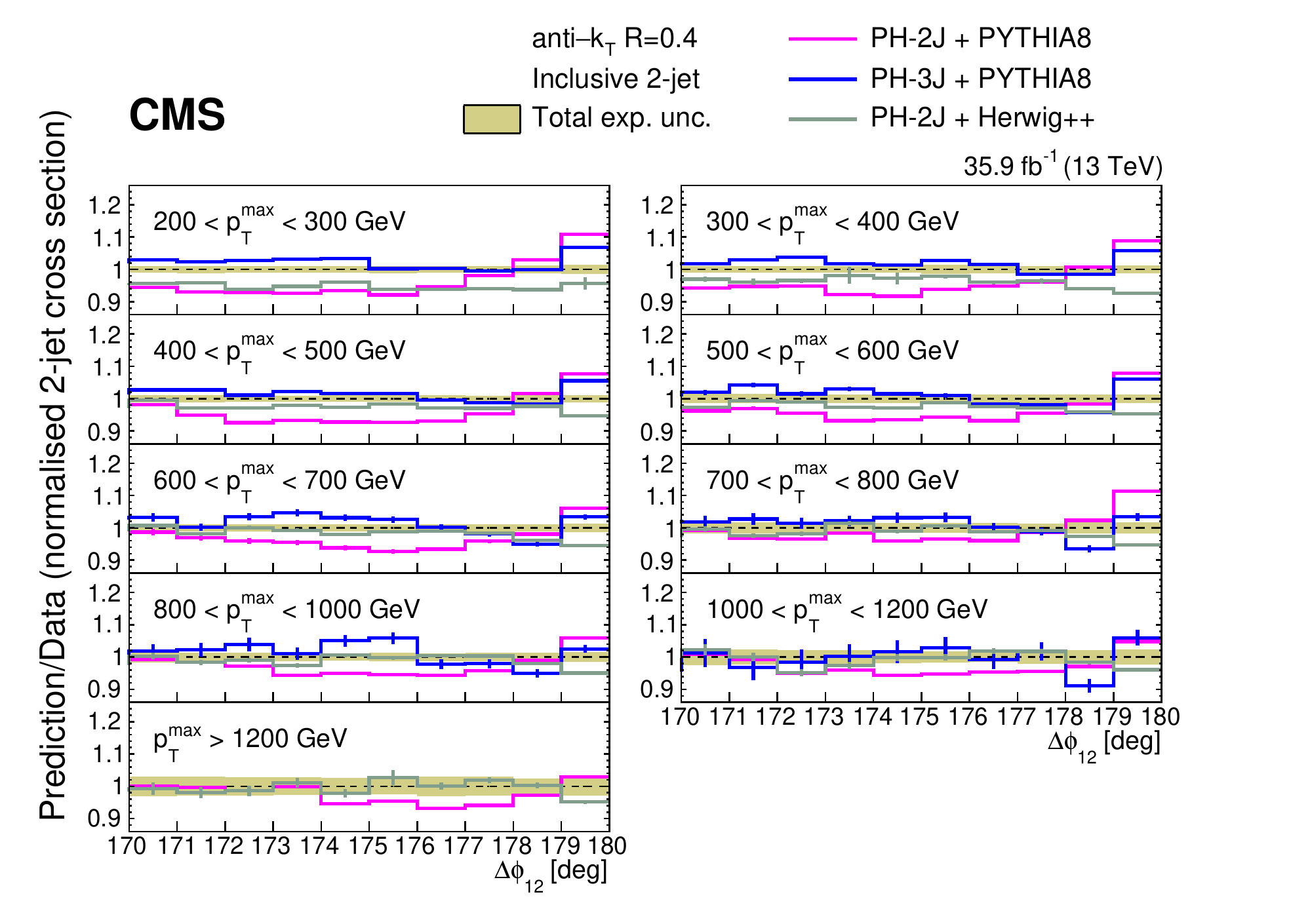}
  \caption{Ratios of the normalized
    inclusive 2-jet distributions for the \POWHEGTWOJET + \PYTHIAE, \POWHEGTHREEJET + \PYTHIAE , and  \POWHEGTWOJET + \HERWIGpp predictions to data  as a function
    of the azimuthal   separation of the two leading jets  \dphiOneTwo, for all the \ptmax regions.
    The solid band indicates the total experimental uncertainty and
    the error bars on the MC points represent the statistical
    uncertainty of the simulated data. The
    \POWHEGTHREEJET prediction is not shown for the highest bin in \ptmax\ because of the large statistical fluctuations.}
  \label{fig:2J_NLO_data}
\end{figure*}

The ratios of the NLO predictions to data for the unfolded normalized inclusive 2-jet distributions for the different \ptmax regions are shown in Fig.~\ref{fig:2J_NLO_data}.
The NLO calculations considered are \POWHEGTWOJET + \PYTHIAE, \POWHEGTWOJET + \HERWIGpp, and \POWHEGTHREEJET + \PYTHIAE.
Among these NLO predictions  \POWHEGTHREEJET + \PYTHIAE agrees better with the data. The \POWHEGTWOJET + \HERWIGpp prediction is similar to the one of \POWHEGTHREEJET + \PYTHIAE , except for the lowest \ptmax region.

\begin{figure*}[htb]
  \centering
  \includegraphics[width=0.4\textwidth]{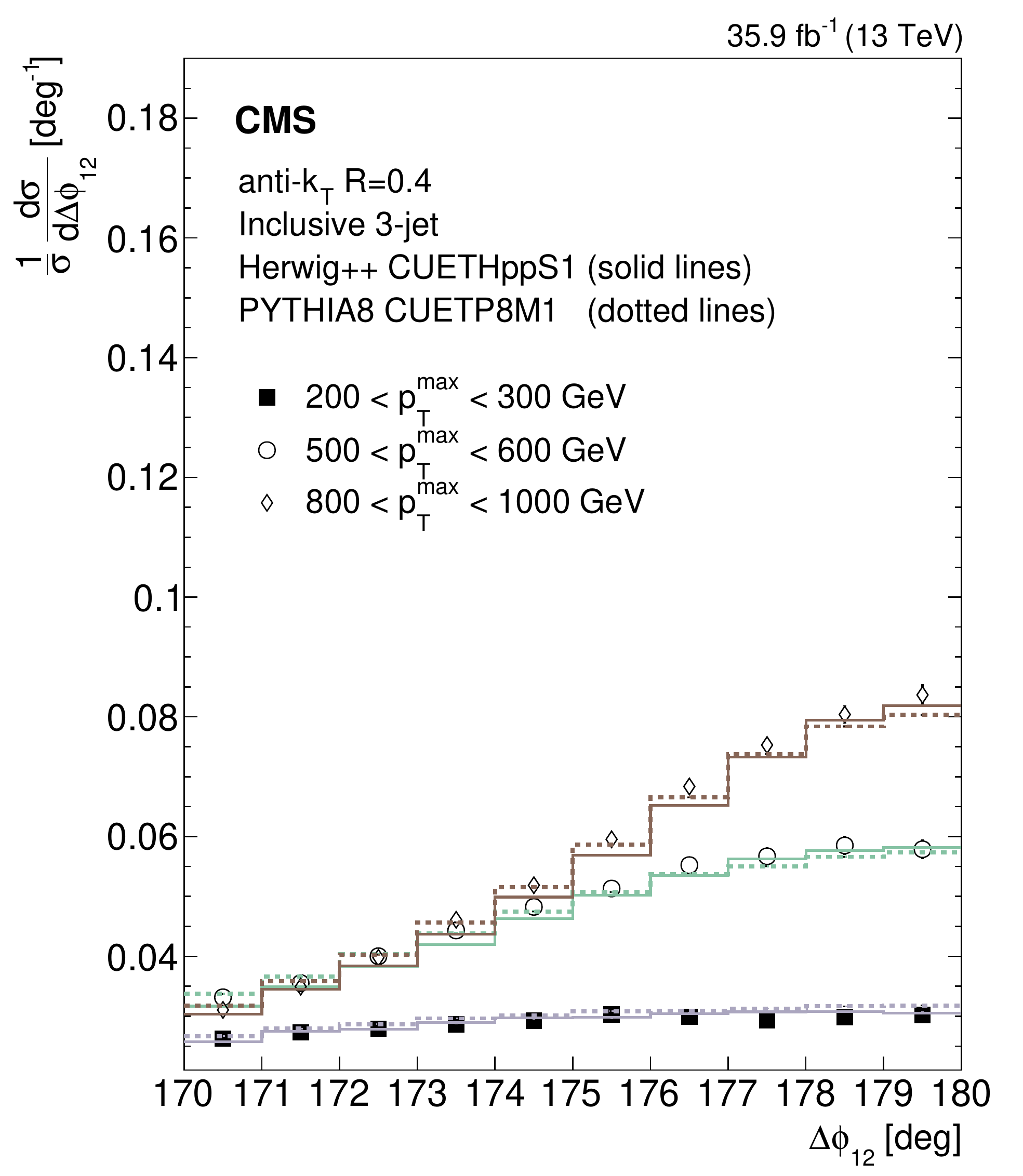}
  \includegraphics[width=0.4\textwidth]{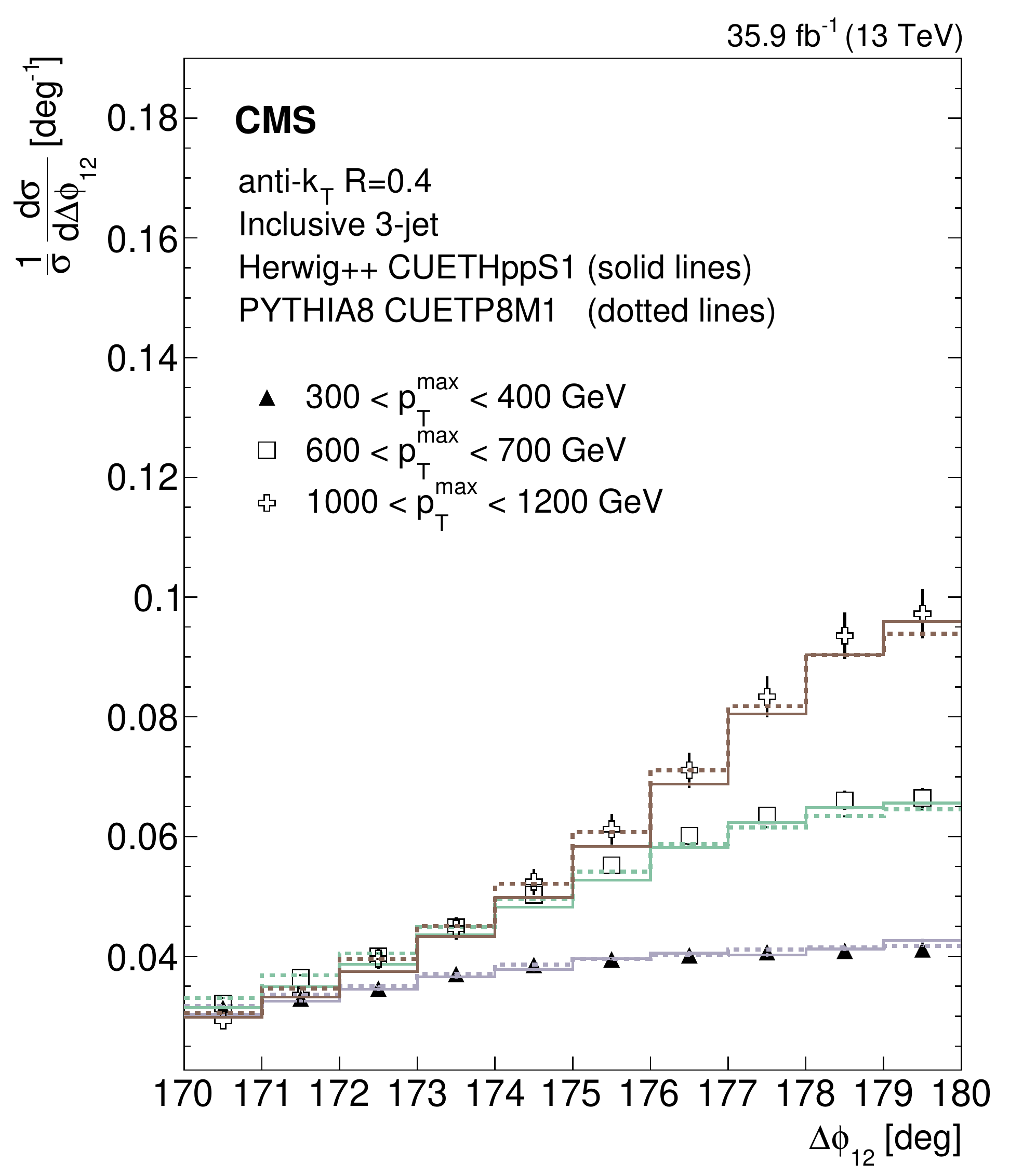}
  \caption{Normalized inclusive 3-jet distributions as a function of the azimuthal separation of the two leading jets \dphiOneTwo
    for different \ptmax regions.
 The data are represented by the markers and the theory by histograms. Overlaid with the data are predictions from the \HERWIGpp event generator (solid lines) and \PYTHIAE (dotted lines). The total experimental uncertainty is depicted as error bars on the measurement.}
  \label{fig:3J_particle_xsection_MC_data}
\end{figure*}

\begin{figure*}[hbt!]
  \centering
  \includegraphics[width=0.8\textwidth]{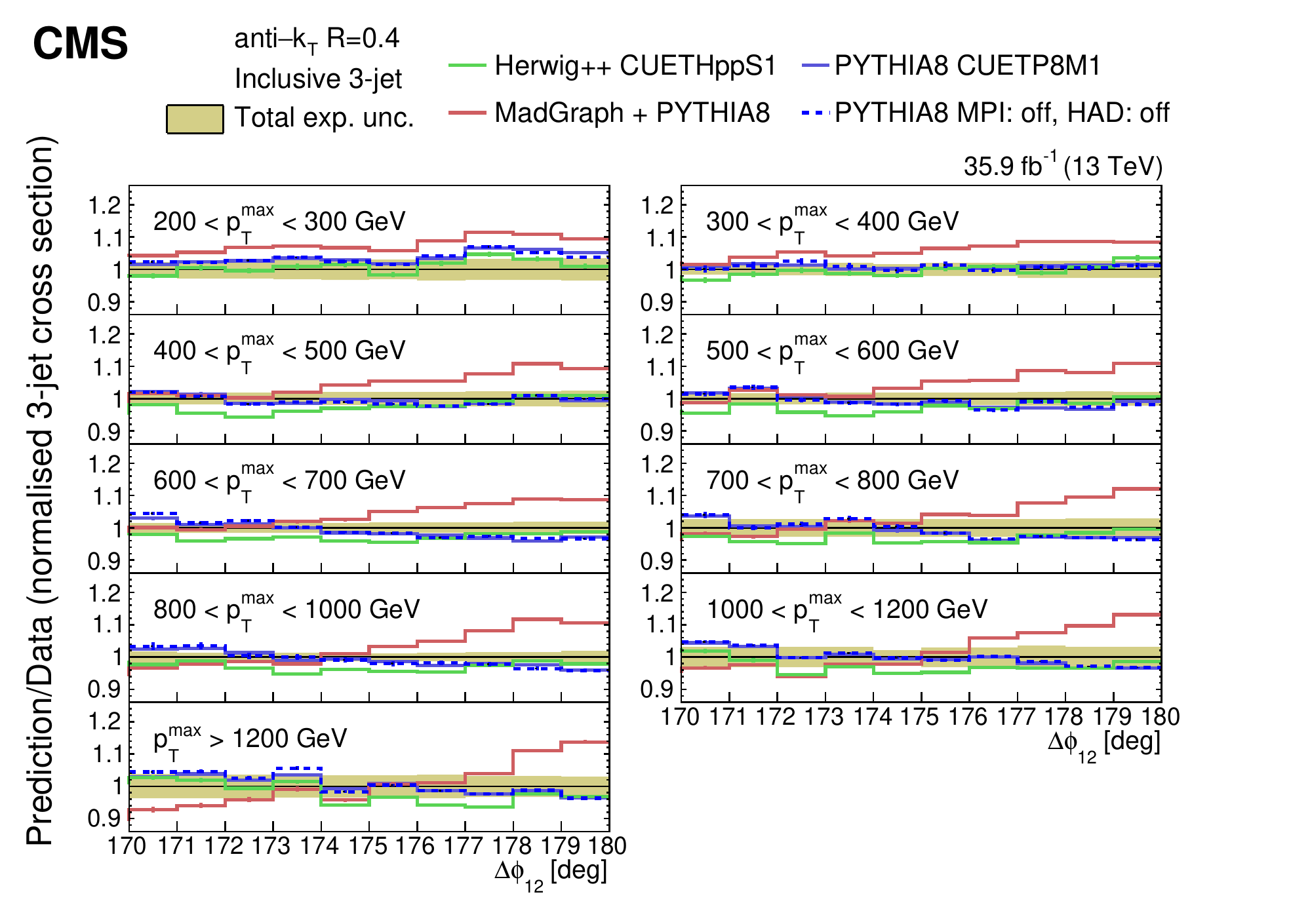}
  \caption{Ratios of the normalized
    inclusive 3-jet distributions for the \PYTHIAE, \HERWIGpp, and \MADGRAPH + \PYTHIAE predictions to data  as a function of the azimuthal   separation of the two leading jets \dphiOneTwo, for all the \ptmax regions.
    The solid band indicates the total experimental uncertainty and
    the error bars on the MC points represent the statistical
    uncertainty of the simulated data.}
  \label{fig:3J_LO_data}
\end{figure*}

In Fig.~\ref{fig:3J_particle_xsection_MC_data} the unfolded normalized inclusive 3-jet distribution as a function of \dphiOneTwo are compared with the predictions from \HERWIGpp  (solid lines) and \PYTHIAE (dotted lines) for different  \ptmax regions.
The ratios of the normalized inclusive 3-jet distributions  for the  \PYTHIAE, \HERWIGpp, and \MADGRAPH + \PYTHIAE
predictions to data are shown in Fig.~\ref{fig:3J_LO_data} for the different \ptmax regions. In contrast to the 2-jet case, \MADGRAPH + \PYTHIAE shows the largest deviations from the measurements close to $180^{\circ}$, whereas \PYTHIAE and \HERWIGpp give a good description of the data.

\begin{figure*}[hbt!]
  \centering
  \includegraphics[width=0.8\textwidth]{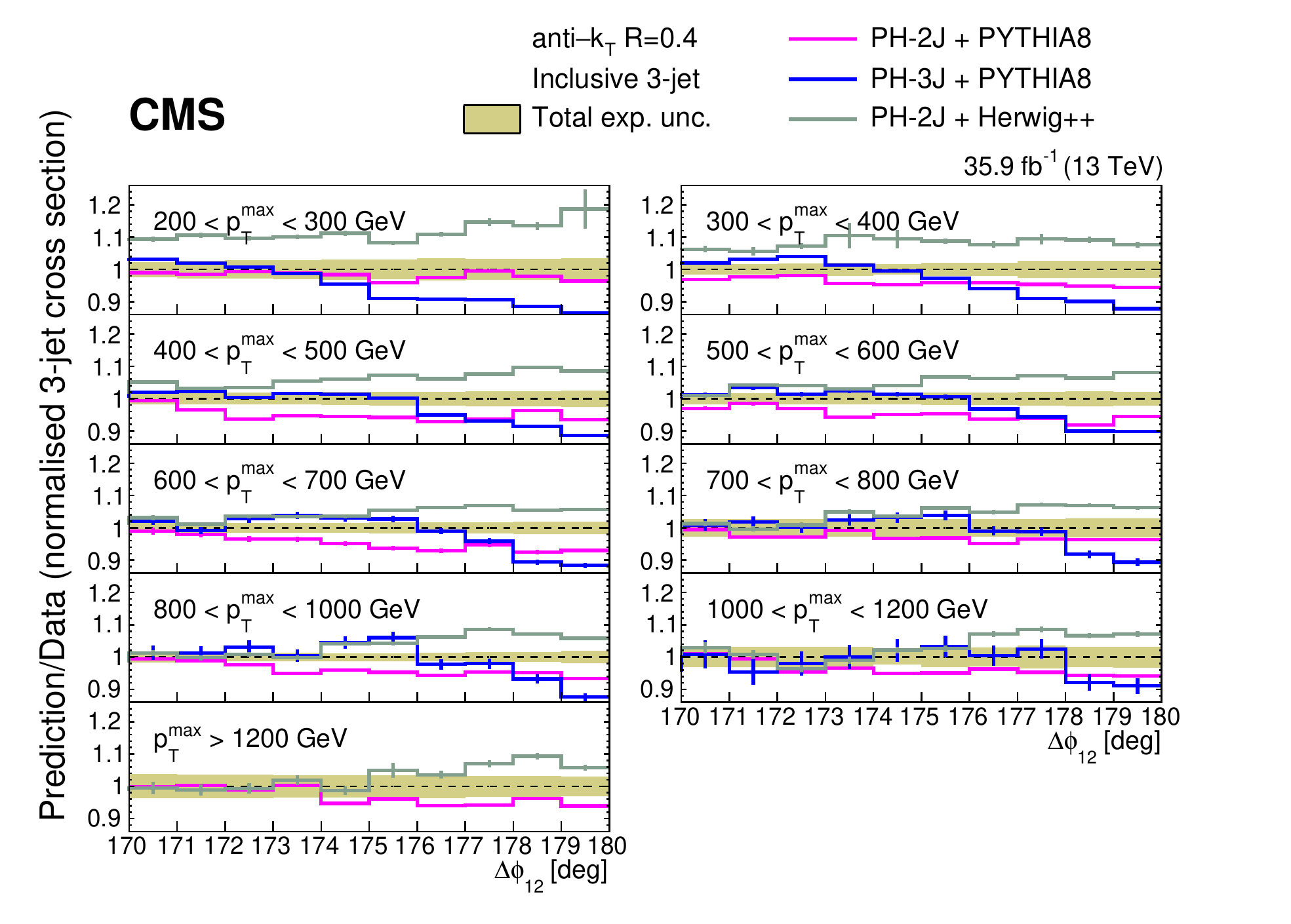}
  \caption{Ratios of the normalized
    inclusive 3-jet distributions for the \POWHEGTWOJET + \PYTHIAE,
    \POWHEGTHREEJET + \PYTHIAE, and \POWHEGTWOJET + \HERWIGpp\  predictions to data  as a function
    of the azimuthal   separation of the two leading jets \dphiOneTwo, for all \ptmax regions.
    The solid band indicates the total experimental uncertainty and
    the error bars on the MC points represent the statistical
    uncertainty of the simulated data. The \POWHEGTHREEJET prediction is not shown for the highest bin in \ptmax\ because of the large statistical fluctuations.}
  \label{fig:3J_NLO_data}
\end{figure*}

\begin{tolerant}{1200}
The ratios of the NLO predictions from \POWHEGTWOJET + \PYTHIAE, \POWHEGTWOJET + \HERWIGpp, and \POWHEGTHREEJET + \PYTHIAE
to data for the normalized inclusive 3-jet distributions are shown in Fig.~\ref{fig:3J_NLO_data}.
 All the considered NLO+PS predictions fail to describe the measurements close to $180^\circ$.
The predictions from \POWHEGTHREEJET and \MADGRAPH (Fig.~\ref{fig:3J_LO_data}) behave very differently, in contrast  to their similar trend in the inclusive 2-jet case.
\end{tolerant}

Since \PYTHIAE, \POWHEGTWOJET + \PYTHIAE, \POWHEGTHREEJET + \PYTHIAE, and \MADGRAPH + \PYTHIAE use the same parton shower, the observed differences in the predictions can be attributed to the treatment of the additional partons present in the \POWHEG and \MADGRAPH ME.

In general we observe that the \dphiOneTwo region close to $180^\circ$ is not well described by the predictions. The predictions agree better with the measurements for increasing \ptmax and moving further away from the back-to-back region in  \dphiOneTwo , where the contribution of resummation effects becomes smaller \cite{Sun:2014gfa}.
The fact that none of the generators is able to describe the 2- and 3-jet measurements simultaneously suggests that the observed differences (of the order of 10\%) are related to the way soft partons are simulated within the PS. The observed differences between \pt and angular ordered PS for the LO generators \PYTHIAE\ and \HERWIGpp are small (Figs.~\ref{fig:2J_LO_data} and~\ref{fig:3J_LO_data}) compared to the \MADGRAPH predictions, which can be attributed to the presence of higher order ME.

\begin{figure*}[hbt!]
  \centering
  \includegraphics[width=0.8\textwidth]{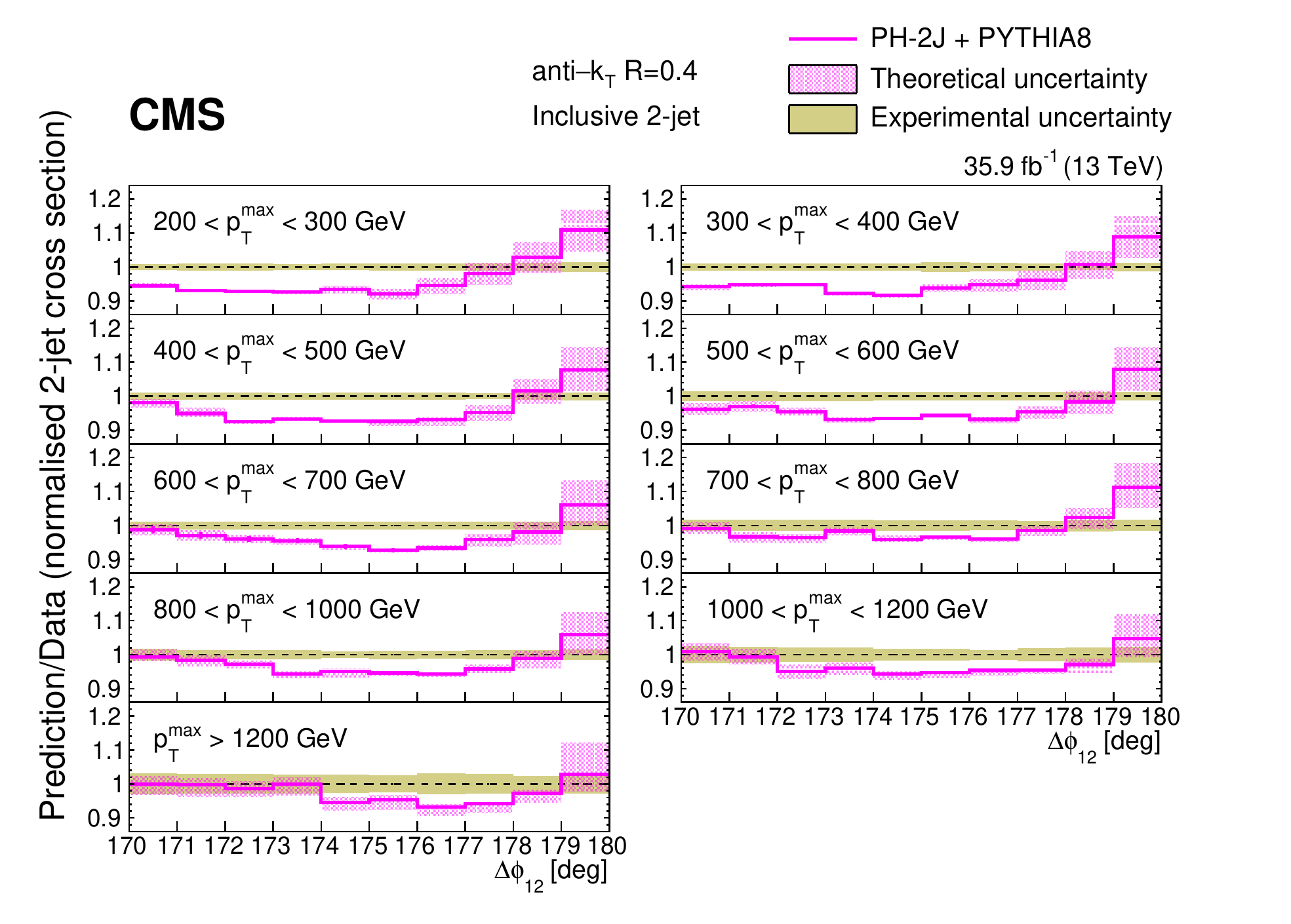}
  \caption{Ratios of the normalized
    inclusive 2-jet distributions for the \POWHEGTWOJET + \PYTHIAE
    predictions to data  as a function  of the azimuthal   separation of the two leading jets  \dphiOneTwo, for all \ptmax regions.
    The solid beige band indicates the total experimental uncertainty and
    the hatched band represents the total theoretical uncertainty.}
  \label{fig:2J_NLO_th}
\end{figure*}

\begin{figure*}[hbt!]
  \centering
  \includegraphics[width=0.8\textwidth]{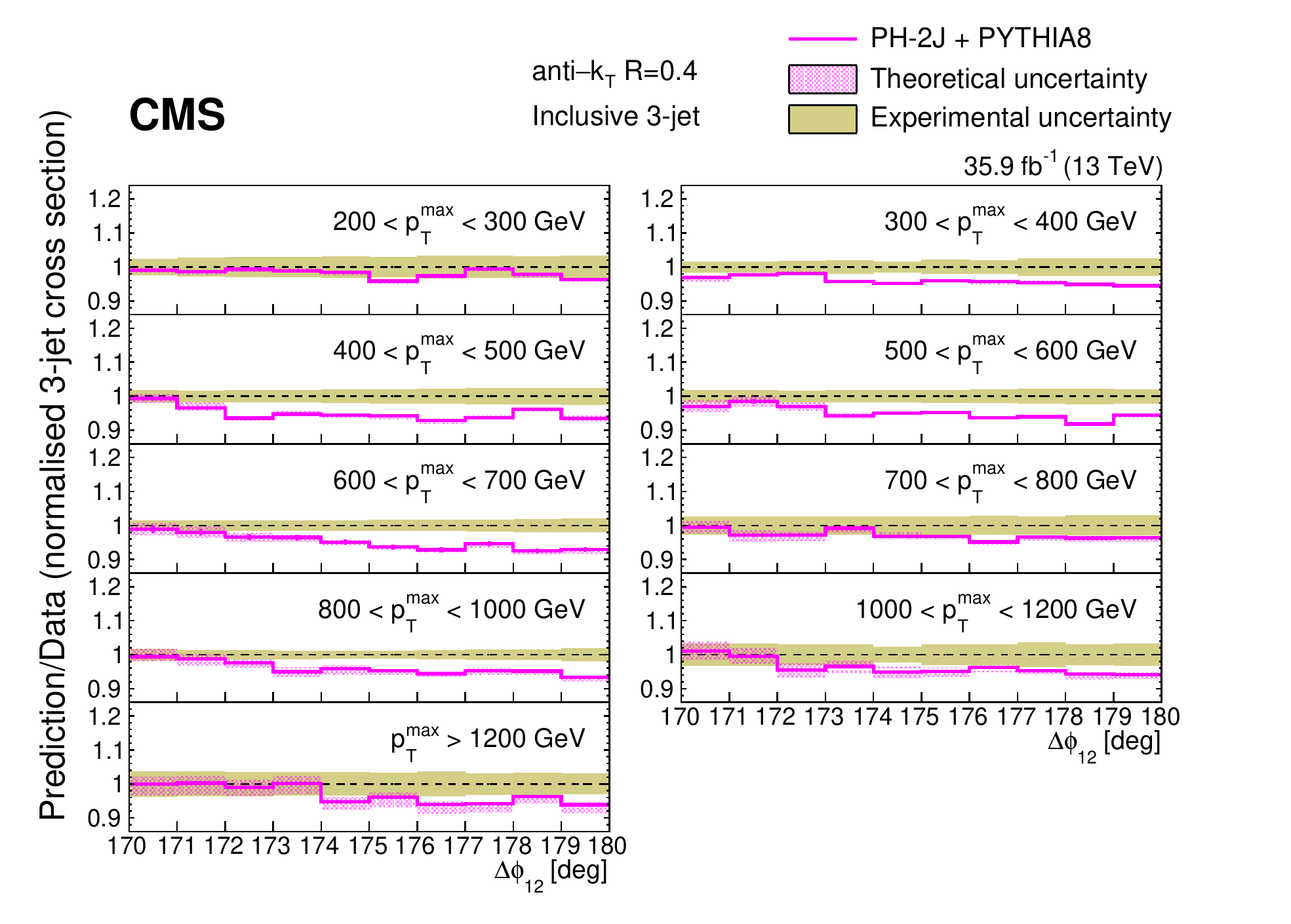}
  \caption{Ratios of the normalized
    inclusive 3-jet distributions for the \POWHEGTWOJET + \PYTHIAE
    predictions to data  as a function of the azimuthal   separation of the two leading jets \dphiOneTwo, for all \ptmax regions.
    The solid beige band indicates the total experimental uncertainty,
    the hatched band represents the total theoretical uncertainty.}
  \label{fig:3J_NLO_th}
\end{figure*}

The theoretical calculations have an intrinsic uncertainty arising from the freedom of choice of the renormalization and factorization scales (\mur and \muf), the choice of the PDF and \alpsmz, and the modeling of nonperturbative effects and PS. The total theoretical uncertainty is the quadratic sum of the uncertainties from the scale, PDF, \alpS, and PS variations. Despite the better agreement of \POWHEGTHREEJET, the \POWHEGTWOJET event generator is used instead for the estimation of the scale, PDF, and \alpS\ uncertainties, because of the larger event sample. For the estimation of the PS uncertainty \PYTHIAE is utilized. The following  four sources of theoretical uncertainties are analyzed:
\begin{itemize}
\begin{tolerant}{1200}
\item The uncertainties due to the renormalization and factorization scales of the hard process are evaluated by varying the default choice $\mur = \muf = \pt$ of the underlying Born configuration
between \pt/2 and 2\pt.
The envelope of the following seven combinations is considered: $(\mur/\pt , \muf/\pt) = (0.5, 0.5)$, $(0.5,1)$, $(1, 0.5)$, $(1,1)$, $(1,2)$, $(2,1)$, and $(2,2)$.
\end{tolerant}
\item The PDF uncertainties are evaluated
according to the prescriptions for the NNPDF3.0 NLO PDF set. There are 100 replicas of the NNPDF3.0 NLO PDF set. For each replica the cross section is calculated and the uncertainty is taken as the envelope from all the replicas.
\item The uncertainty due to the value of the strong coupling \alpS\ is obtained by a variation of \alpsmz  by $\pm$0.001, as recommended in Ref.~\cite{Lai:2010nw}.
\item The uncertainty due to PS is evaluated with the \PYTHIAE\ event generator by varying the default renormalization scale choice $\mur = \pt$ of the branching in initial state (ISR) and  final state radiation (FSR) between $\mur/2$ and 2$\mur$.
The envelope of the following nine combinations is considered: (ISR $\mur/\pt$ , FSR $\mur/\pt$) $= (0.5,0.5)$, $(0.5,1)$, $(0.5,2)$, $(1,0.5)$, $(1,1)$, $(1,2)$, $(2,0.5)$, $(2,1)$, and $(2,2)$.
\end{itemize}

The nonperturbative contributions (MPI and hadronization) are included in the calculations above.
 The uncertainty from these contributions are estimated from the different choices of the UE tune and found to be negligible.

The uncertainty from PS dominates for the normalized inclusive 2-jet distributions. It is one order of magnitude larger than the rest of the sources near $\dphiOneTwo=180^{\circ}$. On the other hand, for the normalized inclusive 3-jet distributions, the main contributions come from PS and PDF uncertainties.
 The predictions from \POWHEGTWOJET + \PYTHIAE and \POWHEGTWOJET + \HERWIGpp (Fig.~\ref{fig:2J_NLO_data}) show the differences from using different PS models together with different matching procedures.

Figs.~\ref{fig:2J_NLO_th}(\ref{fig:3J_NLO_th}) show the ratios of the \POWHEGTWOJET predictions to data for the normalized inclusive 2(3)-jet distributions for the different \ptmax  regions.
The solid beige band indicates the total experimental uncertainty, and
the hatched band represents the total theoretical uncertainty.

For the inclusive 2-jet distributions, the theoretical uncertainty is larger than the experimental one in the region close to $\dphiOneTwo=180^{\circ}$ (Fig.~\ref{fig:2J_NLO_th}). This is because the contribution from PS dominates in this region, and its uncertainty is large. For the inclusive 3-jet distributions (Fig.~\ref{fig:3J_NLO_th}), the theoretical uncertainty is smaller in the region close to $180^\circ$. In this case, the region close to $180^\circ$ is not filled by the partons from the PS, but by the third parton from \POWHEGTWOJET, leading to a smaller PS uncertainty.

\section{Summary}
\label{sec:summary}

Measurements of the normalized inclusive 2- and 3-jet distributions as a function of the azimuthal separation \dphiOneTwo between the two jets with the highest transverse momentum \pt , in the collinear back-to-back region, are
presented for several \ptmax\  ranges of the leading jet.
The measurements are performed using data collected  with the CMS experiment at the LHC,
corresponding to an integrated luminosity of 35.9\fbinv of \Pp\Pp\ collisions at a center-of-mass energy of $13\TeV$.

The measured \dphiOneTwo distributions generally agree with predictions from \PYTHIAE, \HERWIGpp,
\MADGRAPH + \PYTHIAE, \POWHEGTWOJET + \HERWIGpp,  and \POWHEG (\POWHEGTWOJET and \POWHEGTHREEJET)  matched to \PYTHIAE.
Discrepancies between the measurement and theoretical predictions are as large as 15\%, mainly in the region $177^\circ < \dphiOneTwo < 180^\circ$.
The predictions agree better with the measurements for larger \ptmax and smaller \dphiOneTwo , where the contribution of resummation effects becomes smaller.
The 2- and 3-jet measurements are not simultaneously described by any of models.

The tree-level multijet event generator \MADGRAPH in combination with \PYTHIAE for showering,
hadronization, and multiparton interactions, shows deviations from the measured \dphiOneTwo for the inclusive
2-jet case, and even larger deviations for the 3-jet case.
The \PYTHIAE and \HERWIGpp predictions show deviations (up to 10\%) for the 2-jet inclusive distributions, whereas their  predictions are in reasonable agreement with the inclusive 3-jet distributions.

The next-to-leading-order \POWHEGTWOJET + \PYTHIAE prediction does not describe the data and  a different trend compared to  \PYTHIAE\ and \HERWIGpp\  towards $\dphiOneTwo=180^{\circ}$ is observed. The \POWHEGTHREEJET + \PYTHIAE predictions agree with the measurements except for the last bin in the low \ptmax intervals. The \POWHEGTWOJET + \HERWIGpp prediction agrees well with the measurement in the highest \ptmax ranges.
For the inclusive 3-jet case, \POWHEGTWOJET + \PYTHIAE performs similarly to \PYTHIAE and \HERWIGpp in the whole \dphiOneTwo range for high \ptmax intervals. \MADGRAPH + \PYTHIAE, \POWHEGTHREEJET + \PYTHIAE, and \POWHEGTWOJET + \HERWIGpp show deviations from the measurements of up to 15\%.

The measurement of correlations for collinear back--to--back dijet configurations probes the multiple scales involved in the event and, therefore, the differences observed between predictions and the measurements illustrate the importance of improving the models of soft parton radiation accompanying the hard process.

 \begin{acknowledgments}
We congratulate our colleagues in the CERN accelerator departments for the excellent performance of the LHC and thank the technical and administrative staffs at CERN and at other CMS institutes for their contributions to the success of the CMS effort. In addition, we gratefully acknowledge the computing centers and personnel of the Worldwide LHC Computing Grid for delivering so effectively the computing infrastructure essential to our analyses. Finally, we acknowledge the enduring support for the construction and operation of the LHC and the CMS detector provided by the following funding agencies: BMBWF and FWF (Austria); FNRS and FWO (Belgium); CNPq, CAPES, FAPERJ, FAPERGS, and FAPESP (Brazil); MES (Bulgaria); CERN; CAS, MoST, and NSFC (China); COLCIENCIAS (Colombia); MSES and CSF (Croatia); RPF (Cyprus); SENESCYT (Ecuador); MoER, ERC IUT, and ERDF (Estonia); Academy of Finland, MEC, and HIP (Finland); CEA and CNRS/IN2P3 (France); BMBF, DFG, and HGF (Germany); GSRT (Greece); NKFIA (Hungary); DAE and DST (India); IPM (Iran); SFI (Ireland); INFN (Italy); MSIP and NRF (Republic of Korea); MES (Latvia); LAS (Lithuania); MOE and UM (Malaysia); BUAP, CINVESTAV, CONACYT, LNS, SEP, and UASLP-FAI (Mexico); MOS (Montenegro); MBIE (New Zealand); PAEC (Pakistan); MSHE and NSC (Poland); FCT (Portugal); JINR (Dubna); MON, RosAtom, RAS, RFBR, and NRC KI (Russia); MESTD (Serbia); SEIDI, CPAN, PCTI, and FEDER (Spain); MOSTR (Sri Lanka); Swiss Funding Agencies (Switzerland); MST (Taipei); ThEPCenter, IPST, STAR, and NSTDA (Thailand); TUBITAK and TAEK (Turkey); NASU and SFFR (Ukraine); STFC (United Kingdom); DOE and NSF (USA).

\hyphenation{Rachada-pisek} Individuals have received support from the Marie-Curie program and the European Research Council and Horizon 2020 Grant, contract No. 675440 (European Union); the Leventis Foundation; the A.P.\ Sloan Foundation; the Alexander von Humboldt Foundation; the Belgian Federal Science Policy Office; the Fonds pour la Formation \`a la Recherche dans l'Industrie et dans l'Agriculture (FRIA-Belgium); the Agentschap voor Innovatie door Wetenschap en Technologie (IWT-Belgium); the F.R.S.-FNRS and FWO (Belgium) under the ``Excellence of Science -- EOS" -- be.h project n.\ 30820817; the Beijing Municipal Science \& Technology Commission, No. Z181100004218003; the Ministry of Education, Youth and Sports (MEYS) of the Czech Republic; the Lend\"ulet (``Momentum") Program and the J\'anos Bolyai Research Scholarship of the Hungarian Academy of Sciences, the New National Excellence Program \'UNKP, the NKFIA research grants 123842, 123959, 124845, 124850, and 125105 (Hungary); the Council of Science and Industrial Research, India; the HOMING PLUS program of the Foundation for Polish Science, cofinanced from European Union, Regional Development Fund, the Mobility Plus program of the Ministry of Science and Higher Education, the National Science Center (Poland), contracts Harmonia 2014/14/M/ST2/00428, Opus 2014/13/B/ST2/02543, 2014/15/B/ST2/03998, and 2015/19/B/ST2/02861, Sonata-bis 2012/07/E/ST2/01406; the National Priorities Research Program by Qatar National Research Fund; the Programa Estatal de Fomento de la Investigaci{\'o}n Cient{\'i}fica y T{\'e}cnica de Excelencia Mar\'{\i}a de Maeztu, grant MDM-2015-0509 and the Programa Severo Ochoa del Principado de Asturias; the Thalis and Aristeia programmes cofinanced by EU-ESF and the Greek NSRF; the Rachadapisek Sompot Fund for Postdoctoral Fellowship, Chulalongkorn University and the Chulalongkorn Academic into Its 2nd Century Project Advancement Project (Thailand); the Welch Foundation, contract C-1845; and the Weston Havens Foundation (USA).
 \end{acknowledgments}
\clearpage

\bibliography{auto_generated}

\cleardoublepage \appendix\section{The CMS Collaboration \label{app:collab}}\begin{sloppypar}\hyphenpenalty=5000\widowpenalty=500\clubpenalty=5000\vskip\cmsinstskip
\textbf{Yerevan Physics Institute, Yerevan, Armenia}\\*[0pt]
A.M.~Sirunyan, A.~Tumasyan
\vskip\cmsinstskip
\textbf{Institut f\"{u}r Hochenergiephysik, Wien, Austria}\\*[0pt]
W.~Adam, F.~Ambrogi, E.~Asilar, T.~Bergauer, J.~Brandstetter, E.~Brondolin, M.~Dragicevic, J.~Er\"{o}, A.~Escalante~Del~Valle, M.~Flechl, R.~Fr\"{u}hwirth\cmsAuthorMark{1}, V.M.~Ghete, J.~Hrubec, M.~Jeitler\cmsAuthorMark{1}, N.~Krammer, I.~Kr\"{a}tschmer, D.~Liko, T.~Madlener, I.~Mikulec, N.~Rad, H.~Rohringer, J.~Schieck\cmsAuthorMark{1}, R.~Sch\"{o}fbeck, M.~Spanring, D.~Spitzbart, A.~Taurok, W.~Waltenberger, J.~Wittmann, C.-E.~Wulz\cmsAuthorMark{1}, M.~Zarucki
% ...
\vskip\cmsinstskip
% ...
1:  Also at Vienna University of Technology, Vienna, Austria\\
% ...
\end{sloppypar}
\end{document}